\documentclass[prl,twocolumn,nopacs,superscriptaddress]{revtex4-2}
\usepackage[utf8]{inputenc}
\usepackage[american,]{babel}
\usepackage[T1]{fontenc}
\usepackage[pdftex]{graphicx}  
\usepackage{graphicx, xcolor}
\usepackage{dcolumn}
\usepackage{bm}
\usepackage{dsfont}
\usepackage{amsmath,amsthm,amssymb}
\usepackage{hyperref}
\usepackage[T1,T2A]{fontenc}
\usepackage{xcolor}
\hypersetup{colorlinks,bookmarksopen,bookmarksnumbered,
    citecolor=blue,
    linkcolor=blue,
    pdfstartview=false,
    urlcolor=blue}
\usepackage{graphicx}
\usepackage{braket}
\usepackage{wrapfig}
\usepackage{soul}
\usepackage{mathtools}
\usepackage{mathrsfs}
\usepackage{float}
\usepackage{tabularx}
\usepackage{physics}

\newcommand{\der}[2]{\partial_{#1} #2}
\usepackage{tikz}
\usetikzlibrary{arrows.meta}
\usetikzlibrary{decorations.pathmorphing}

\begin{document}

\title{Entanglement Dynamics across a Monitored Quantum Point Contact}

\author{Anna Delmonte}
\affiliation{JEIP, UAR 3573 CNRS, Coll\`ege de France, PSL Research University, 75321 Paris Cedex 05, France}

\author{Marco Schir\`o}
\affiliation{JEIP, UAR 3573 CNRS, Coll\`ege de France, PSL Research University, 75321 Paris Cedex 05, France}

\date{\today} 

\begin{abstract}
We compute the entanglement dynamics across a monitored quantum point contact, where particle losses are recorded on a given site, and demonstrate how this single-site local monitoring substantially reshapes the entanglement production. Contrarily to the unitary case, where entanglement entropy grows logarithmically in time, here we find first a linear growth, up to a maximum value displaying volume-law scaling, and then a slow decay to zero, as the system empties out. We capture this crossover using a quasiparticle picture, where the first linear growth arises due to an emergent bias voltage established by the losses, which eventually decays away as the system depletes. We connect our results to studies of the Page curve and to experimentally relevant probes, via full counting statistics of charge transfer across a subregion, with only a single channel to unravel leading to a favorable scaling of the postselection overhead. Natural platforms for this setting include mesoscopic systems and ultracold atoms.
\end{abstract}

\maketitle

\label{sec:intro}

\emph{Introduction --} A classic result in the theory of entanglement of quantum matter is that opening and closing a quantum point contact (QPC) leads to an entanglement entropy growing logarithmically in time~\cite{klich2009quantum,song2011entanglement,song2012bipartite,thomas2015entanglement} due to gapless electron-hole excitations~\cite{beenakker2003proposal,beenakker2004quantum,beenakkerc.w.j._2006}, a specific example of a broader class of local quantum quenches  giving rise to critical entanglement growth~\cite{calabrese2005evolution,eisler2012entanglement,stephan2011local,calabrese2016quantum}. Recent years have demonstrated that qualitatively new entanglement patterns can emerge in the quantum trajectories of continuously monitored quantum systems~\cite{wiseman2009quantum,fazio2025manybodyopenquantumsystems}, including for example  measurement-induced phase transitions~\cite{li2018quantum,li2019measurement,skinner2019measurement,cao2019entanglementina,fuji2020measurementinducedquantum,vanregemortel2021entanglement,
turkeshi2021measurementinducedentanglement,coppola2022growthofentanglement,poboiko2023theory,legal2024entanglement,soares2025entanglement} or measurement-altered criticality~\cite{murciano2023measurementaltered,garratt2023measurements,Paviglianiti2024enhanced,naus2025practicalroadmapmeasurementalteredcriticality}. Although intriguing, this effect remains largely theoretical since experimental demonstration requires post-selection~\cite{noel2021measurementinducedquantum,koh2022experimentalrealizationof,hoke2023quantuminformationphases}, which is a priori exponentially costly in the number of measurements in space-time, although several shortcuts have been proposed~\cite{ippoliti2021postselection,li2023cross,passarelli2024many,delmonte2025measurement}. 

Monitored quantum transport offers a natural and experimentally viable platform where to study the non-trivial effects of quantum measurements on entanglement and full counting statistics. Recent experiments with ultra cold gases, for example, have succeeded realizing atomic point contacts~\cite{husmann2015connecting}, where dissipation in the form of atom losses have been implemented~\cite{barontini2013controlling,labouvie2016bistability,corman2019quantized,lebrat2019quantized}. In solid-state settings, monitoring charge fluctuations in quantum dots and quantum point contacts is also a natural perspective~\cite{landi2024current,bayer2025realtime}.

On the theoretical front the effect of dissipation and monitoring on quantum transport has attracted fresh new interest~\cite{Tilloy_2014,biele2017controlling,damanet2019controlling,froml2019fluctuation,froml2020ultracold,alba2022noninteracting,alba2022unbounded,
visuri2022symmetry,wampler2022stirring,khor2023measurement,ferreira2023exact,turkeshi2024density,gievers2024quantum,stefanini2024dissipative,vanhoecke2025kondozeno,di2024entanglement,Beenakker2025monitoredquantum}.

In this Letter, we study the entanglement dynamics across a monitored QPC, where at the QPC site, a charge detector is placed and clicks recorded, corresponding to particle loss events. We show that a local monitoring completely restructures the entanglement dynamics, which now displays three regimes: (i) a robust linear growth in time with volume-law scaling at the maximum, (ii) a universal power-law decay and (iii) an exponential tail at long times, corresponding to the full depletion of the system. We explain the three dynamical regimes by the interplay between entanglement and charge dynamics, which we formalize within a quasiparticle picture. To connect our findings to experimentally relevant probes we compute and discuss the behavior of charge cumulants and full-counting statistics (FCS)~\cite{PismaZhETF.58.225,nazarov2009quantum,reulet2003environmental,sukhorukov2007conditional,dasenbrook2016dynamical}, 
which for our setting is directly related to the entanglement entropy~\cite{klich2009quantum,song2011entanglement,song2012bipartite,Calabrese_2012,thomas2015entanglement,poboiko2023theory}. Finally, we connect our findings to studies of the Page curve~\cite{page1993information,page2013time}, recently extended from black-hole physics to isolated and open quantum lattice models ~\cite{kehrein2024page,saha2024generalized,jha2025page,li2025sharp,ganguly2025quantum}.
\begin{figure}
\centering 
\includegraphics[width=0.99\linewidth]{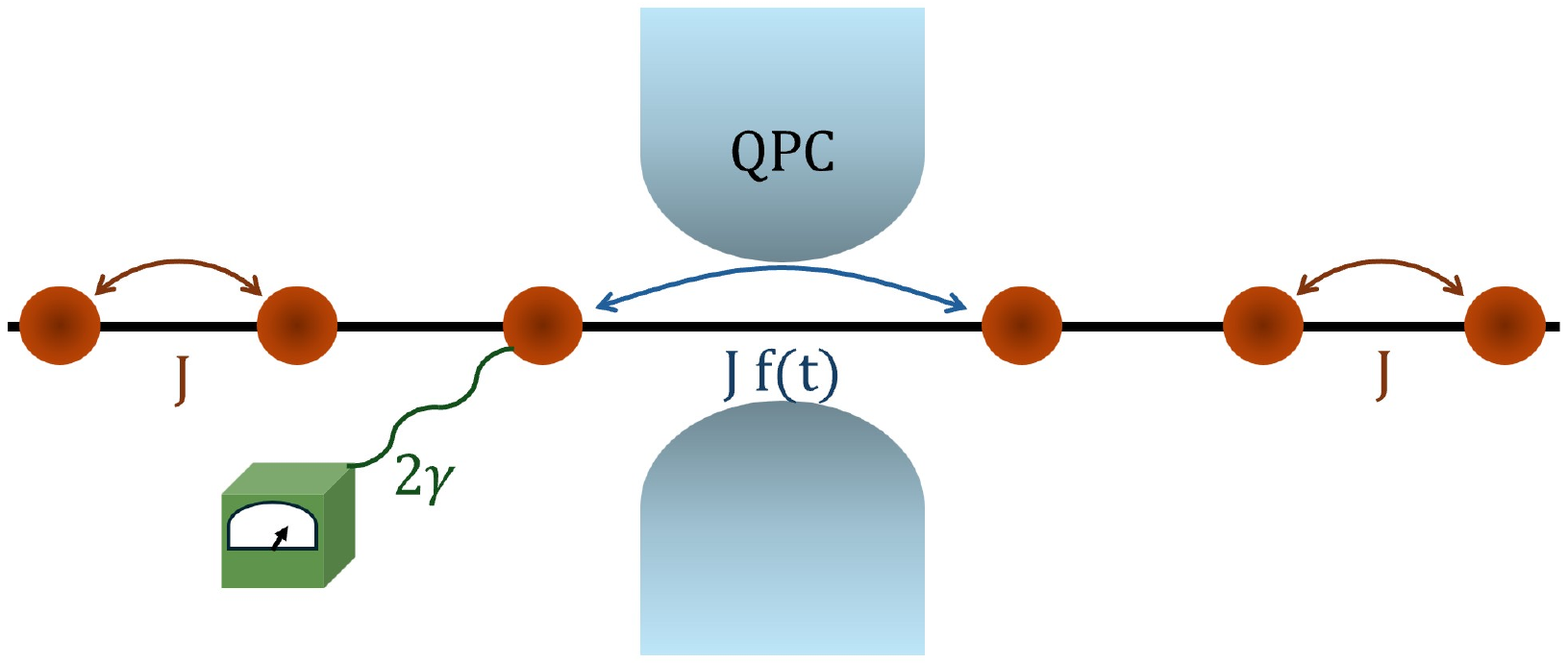}
    \caption{Sketch of the system: two leads described by a tight-binding chain with hopping parameter $J$ are connected via a quantum point contact opening according to $f(t)$. A detector measures particle losses at the last site of the left lead with a rate $2\gamma$.}
    \label{fig:sketch}
\end{figure}
Our setup provides a framework where non-trivial measurement-induced effects on the entanglement entropy between the two leads, such as the production of volume-law scaling through local measurements, can be observed via the full counting statistics of the transported charge. This offers a simplified approach to the challenge of post-selection.

\begin{figure*} 
\centering 
\includegraphics[width=0.99\textwidth]{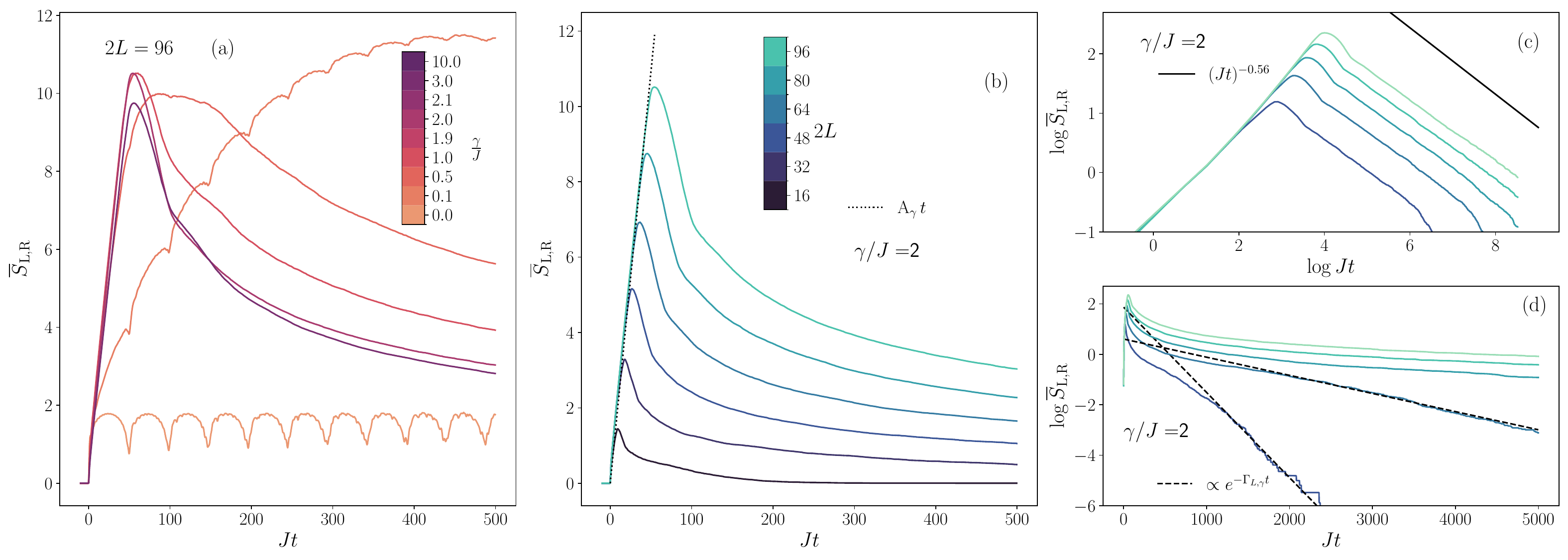} 
    \caption{Dynamics of the entanglement entropy between left and right lead. (a) Dynamics of the entanglement entropy for a chanin of $2L=96$ sites and several measurement strengths. The orange line corresponds to the unitary case $\gamma/J=0$. (b) Comparison of the dynamics for different system sizes at $\gamma/J=2$. The peak of the entanglement entropy scales linearly with the system size. The dotted line fits the initial linear regime.  (c) Polynomial regime of the entanglement entropy at $\gamma/J=2$, the solid black line represents a decay with a universal power $(Jt)^{-0.56}$. The plot is in log-log scale. (d) Exponential regime at long times. The dashed lines represent the exponential fit of the curves for lower system sizes. The plot is in log-linear scale.}
    \label{fig:oms_ent}
\end{figure*} 
 
\emph{Model and Monitoring Protocol -- }We consider a model for a dissipative QPC, connecting two ($\alpha=\rm{L},\rm{R}$) metallic leads, modeled as fermionic tight-binding chains of length $L$ and open boundary conditions. The total Hamiltonian of the system reads $\hat H_0 = \hat H_{\rm{L}} + \hat H_{\rm{R}} + \hat H_{\rm{qpc}}$, where~\cite{thomas2014waiting,thomas2015entanglement} 
\begin{align}
    &\hat H_{\rm{L}} + \hat H_{\rm{R}} = -J\sum_{i=1}^L\,\sum_{\alpha=\rm{L},\rm{R}} \hat c^\dag_{i\alpha}\hat c_{i+1\alpha}+ \, \rm{h.c.} \nonumber\\
    &\hat H_{\rm{qpc}} = -Jf(t) \, \left(\hat c^\dag_{L\rm{L}}\hat c_{1\rm{R}}+\hat c^\dag_{1\rm{R}}\hat c_{L\rm{L}}\right) \nonumber
\end{align}
Here $J$ is the hopping strength and $f(t)=1/2+\arctan(t/\tau)/\pi$ is the opening function of the QPC with opening time $\tau$. We consider a monitoring quantum jump protocol that detects particle loss at the QPC. The model is sketched in Fig.~\ref{fig:sketch}. The system's state evolves according to a stochastic Schrodinger equation~\cite{molmer1993monte,gardiner2004quantum,wiseman2009quantum,jacobs2014quantum,daley2014quantum}
\begin{align}
         \der{t}{\ket{\Psi_{\xi}}} &= -idt\left\{\hat H - \frac{i}{2}\left(\hat M^{\dagger}M - \langle \hat M^{\dagger}\hat M\rangle\right)\right\}\ket{\Psi_{\xi}}\nonumber\\
         &+ d\xi\left\{ \frac{\hat M}{\sqrt{\langle \hat M^{\dagger}\hat M \rangle}} - 1\right\}\ket{\Psi_{\xi}}  \label{eq:sto_schrodi}
\end{align}
with $\hat M = \sqrt{2\gamma}\,\hat c_{L\mathrm{L}}$, the jump operator describing the loss of an electron at the QPC site $i=L$ on the left lead, and $d\xi(t) \in \{0,1\}$ is a state-dependent Poisson increment such that $P(d\xi(t) = 1 ) = dt \bra{\Psi(t)}  \hat M^\dagger \hat M\ket{\Psi(t)} $. The monitored dynamics in Eq.~\eqref{eq:sto_schrodi} can be seen as a piecewise deterministic non-Hermitian evolution, the first line of Eq.~\ref{eq:sto_schrodi}, interrupted by quantum jumps at random times, the second line of Eq.~\ref{eq:sto_schrodi}. 
The dynamics along a quantum trajectory conserves the purity, however averaging the density matrix $\hat \rho_{\xi}=\ket{\Psi_{\xi}}\bra{\Psi_{\xi}}$ over the measurement noise gives rise to a mixed state, $\hat \rho\equiv \overline{\rho_{\xi}}$, which evolves according to the Lindblad equation $\partial_t\hat\rho=-i[\hat H,\hat\rho]+\mathcal D[\hat M]\hat\rho$, with $\mathcal D[\hat M]\bullet = \hat M\bullet \hat M^{\dag}-1/2\{\hat M^\dag\hat M,\bullet\}$. In the following we will denote with brackets the averages over the density matrix $\hat \rho$, i.e. $\langle\bullet \rangle=\mbox{Tr}(\hat \rho\bullet)$. In the present context, this problem describes free fermions with a localized loss and displays slow depletion dynamics which has been studied in detail~\cite{froml2019fluctuation,froml2020ultracold,alba2022noninteracting,alba2022unbounded}. 

Here on the other hand we are interested in the entanglement encoded in the pure-state quantum jump unraveling of this Lindblad master equation, corresponding to Eq.~\eqref{eq:sto_schrodi}. To this extent we compute the entanglement entropy between left and right leads. This is defined as $S_{\rm{L,R}}(\xi,t) = - \mathrm{Tr}_{\rm{L}} \left[ \hat\rho_\xi^{\rm L}(t) \ln \hat\rho_\xi^{\rm L} (t) \right] $ where we have introduced a bi-partition between left and right leads, with the reduced density matrix of the left lead being $\hat\rho^{\rm L}_\xi (t) = \mathrm{Tr}_{\rm{ R}} \lvert \Psi_\xi (t) \rangle \langle \Psi_\xi (t) \rvert $.
Specifically, we will study the evolution of the trajectory average of the entanglement entropy 
\begin{equation}
    \overline{S}_{\rm{L,R}}(t) = \int \mathcal{D} \xi P(\xi) S_{\rm{L,R}} (\xi,t)\,,
    \label{eq:av_entanglement}
\end{equation}
where trajectories are summed according to Born's rule through $P(\xi)$, which represents the probability of measuring the trajectory $\xi$.
We choose, an initial separable state between left and right lead, both being the ground state of the tight-binding Hamiltonians $\hat H_{\mathrm{L,R}}$ at half-filling. We are interested in studying the dynamics of the entanglement entropy from the aforementioned initial state, starting when the quantum point contact is closed. We set the initial time as $Jt_i=-10$ and consider a sudden opening of the quantum point contact with typical opening time $J\tau=10^{-6}$, such that $f(t_i)\sim 0$ as desired.

\emph{Entanglement dynamics -- } \label{sec:ent_discussion}
The dynamics of the entanglement entropy exhibits several interesting properties, arising from the competition between unitary dynamics, responsible for transport and the generation of quantum correlations, and particle loss, which eventually depletes the system and inevitably drives it to a vacuum product state. 
We start discussing how the entanglement dynamics depends on the monitoring rate $\gamma$. A first striking feature emerges in Fig.~\ref{fig:oms_ent}(a): while for  unitary dynamics the entanglement entropy shows the well known logarithmic growth in time~\cite{klich2009quantum,thomas2015entanglement,song2012bipartite} already a small monitoring rate leads to a dramatic entanglement growth. For $\gamma/J=0.1$ for example, we see clear revivals on top of a fast growth which, on the time scale of our simulation, does not reach a steady-state value. Upon further increasing $\gamma/J$ we see a clear maximum appearing in $\overline{S}_{\rm{L,R}}(t)$, followed by a slow decrease in time. Both the time scale at which the maximum entanglement is reached as well as its value at the maximum display a non-monotonic behavior with $\gamma/J$, a signature of a Quantum Zeno Effect (QZE)~\cite{supplemental}.

We then focus on a specific measurement strength $\gamma/J=2$ and discuss more in detail the scaling of the entanglement entropy. In Fig.~\ref{fig:oms_ent}(b) we see that the initial growth is linear in time and reaches a maximum on a time scale that grows with $L$, together with the value of $\overline{S}_{\rm{L,R}}(t)$ at the maximum~\cite{supplemental}. This volume law growth is particularly remarkable and entirely due to the local monitoring process, as we will discuss below. Counter-intuitively, the introduction of local monitoring enhances entanglement, driving the system from the logarithmic scaling of the unitary case~\cite{thomas2015entanglement,eisler2012entanglement,stephan2011local} to a robust volume-law behavior at its maximum. In Fig.~\ref{fig:oms_ent}(c)
we discuss more in detail the decay of the entanglement entropy after its maximum. For intermediate times, we find an algebraic decay $\overline{S}_{\rm{L,R}}(t)\sim 1/t^{\alpha}$   according to a universal power law (solid black line), with $\alpha=0.56$. The time window for this universal decay increases with system size. The behavior at long times is ultimately exponential, as shown in Fig.~\ref{fig:oms_ent}(d), $\overline{S}_{\rm{L,R}}(t)\sim \exp(-\Gamma_{\gamma,L} t)$, with a rate that depends on system size $\Gamma_{\gamma,L}$~\cite{supplemental}.  This decay takes the system to the final vacuum state, which is characterized by vanishing entanglement entropy between left and right leads for all finite sizes.
In the thermodynamic limit, the system does not totally deplete and remains in the first regime we presented.

\begin{figure*} 
\centering 
\includegraphics[width=\textwidth]{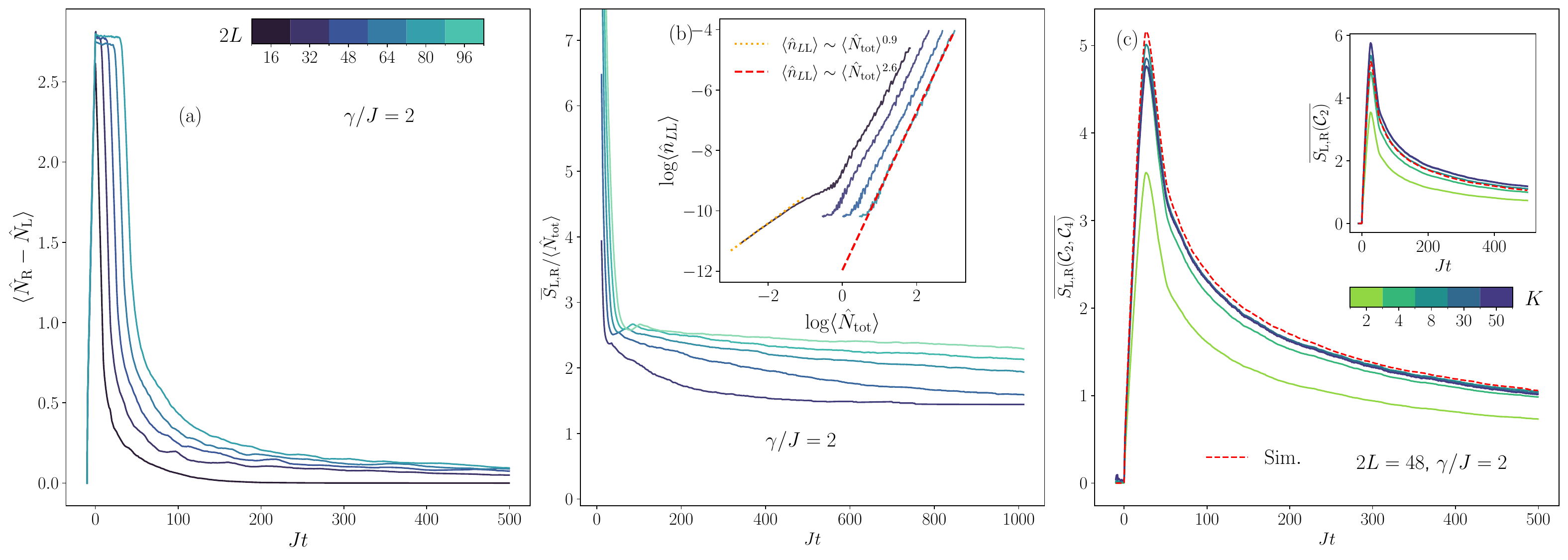} 
    \caption{ (a) Dynamics of the particle imbalance between the two leads for different sizes, and $\gamma/J=2.0$.(b) Ratio between the entanglement entropy and total density for $\gamma/J=2$. In the inset, the total number of particle is plotted as a function of $n_{L\rm{L}}$, and the polynomial and exponential regimes of the entanglement entropy and density emerge respectively from the fits in dashed and dotted lines. (c) Entanglement entropy captured through the cumulants of the transported charge. The main plot considers cumulants $\mathcal C_2,\mathcal C_4$, while the inset considers solely $\mathcal C_2$. Results are shown for several truncations of the cumulant sum, as shown in legend.}
    \label{fig2_qp}
\end{figure*}

\emph{Physical interpretation and Quasiparticle Picture -- } We now provide a complete understanding of the different dynamical regimes of the entanglement entropy and connect them to the charge dynamics across the QPC. First, we note that due to the local dissipation on the left lead, there is a macroscopic particle imbalance that emerges at the QPC, $\langle \hat N_R-\hat N_L\rangle$, as we show for $\gamma/J=2$ and different system sizes in Fig.~\ref{fig2_qp}(a). This imbalance establishes quickly after opening the QPC, remains constant in time up to a time scale that depends on system size, and then drops back to zero. As a result, the QPC is transiently biased by an emergent finite dc voltage $V \sim 4J \sin(\pi(\langle \hat N_{\rm L}\rangle - \langle \hat N_{\rm R}\rangle) / 2L)$, given by the difference in Fermi energies between left and right leads. In the unitary case, a finite bias voltage and a finite transmission probability leads to linear growth in time of the entanglement entropy due to the binomial process of quasiparticles being transmitted across the QPC~\cite{beenakkerc.w.j._2006,nazarov2011flows,thomas2014waiting,klich2009quantum,song2011entanglement,song2012bipartite,thomas2015entanglement}. We can therefore generalize this result to the monitored case: here, scattering of quasiparticles off the monitored QPC leads to finite transmission $T_k$ and reflection $R_k$, even for a perfectly uniform QPC. In addition, the lack of unitarity leads to a finite absorption coefficient $\eta_k$,  such that $T_k+R_k+\eta_k=1$~\cite{froml2019fluctuation,alba2022unbounded,alba2022noninteracting}. We can therefore assume that entanglement production is due to the binomial process of fermion detection/non-detection at the QPC and write the entanglement entropy as $\overline{S}_{\rm{L,R}}(t)=A_\gamma t$ with the slope given by
\begin{align}\label{eqn:slope}
    A_\gamma = \int_{k_{\rm L}}^{k_{\rm R}}\,\frac{dk}{2\pi} \vert v_k\vert {\rm{h}}(1-\eta_k),
\end{align}
with $v_k = -2J \sin(k)$ the group velocity of quasiparticles in the two leads, ${\rm{h}}(x) = -x \log(x) - (1-x) \log(1-x)$ and where $k_{\rm{L}} = \frac{\pi}{L}N_{\rm{L}}^*$, and $k_{\rm R} = \frac{\pi}{L}N_{\rm{R}}^*$ are the maximum filled momenta in the two tight-binding bands on the plateau visible in $\expval{\hat N_{\rm R}-\hat N_{\rm L}}$ in Fig.~\ref{fig2_qp}(a). For a thermodynamically large system $A_\gamma$ reduces to the current flowing through the dissipative QPC~\cite{beenakkerc.w.j._2006,nazarov2011flows,thomas2014waiting,klich2009quantum,song2011entanglement,song2012bipartite,thomas2015entanglement}. As we show in Fig.~\ref{fig:oms_ent}(b) the slope in Eq.~(\ref{eqn:slope}) perfectly matches the numerics for the linear growth of the entanglement entropy.

As long as the bias remains non-zero, the linear growth of the entanglement is visible. When the bias vanishes, the entanglement starts to decay in turn. From Fig.~\ref{fig2_qp}(a) we see indeed that the time duration of the bias coincides with the time interval of the linear growth of the entanglement entropy.
For a finite system, this imbalance cannot be sustained indefinitely and the system eventually starts depleting. The evolution of the total charge $\langle \hat N_{\rm tot}\rangle=\sum_{i\alpha}\langle\hat c^{\dagger}_{i\alpha}\hat c_{i\alpha}\rangle$ is controlled by the Lindblad master equation, which gives $\partial_t \langle \hat N_{\rm{tot}}\rangle = -2\gamma \langle \hat n_{L\rm{L}}\rangle$, i.e. the particle density a the QPC controls the rate of depletion.  Solving the Lindblad dynamics one can show~\cite{supplemental} that the total charge features an initial drop which is linear in time, followed by a universal power-law decay, $\langle \hat N_{\rm{tot}}\rangle\sim 1/(Jt)^{\alpha}$,  since at intermediate times $\langle \hat n_{L\rm{L}}\rangle\propto \langle \hat N_{\rm{tot}}\rangle^{\beta}$, as we show in the inset of Fig.~\ref{fig2_qp}(b) where within our numerical accuracy we have $\beta=(1+\alpha)/\alpha$ and $\alpha=0.56$. At long times,  when the  density profile is homogeneous $\langle \hat n_{L\rm{L}}\rangle\propto \langle \hat N_{\rm{tot}}\rangle/2L$, the depletion dynamics is exponential~\cite{supplemental}.  The long-time decay of the entanglement entropy from its maximum value can be tightly related to this depletion dynamics.  Indeed, as we show in  Fig.~\ref{fig2_qp}(b) that the dynamics of entanglement entropy and total particle's number $\langle \hat N_{\rm tot}\rangle$ are comparable at long times and large system sizes, their ratio approaches a constant value, suggesting their decay is controlled by the same law. 

To support quantitatively the connection between the entanglement entropy and the total number of particles, we now introduce a quasiparticle picture for the average entanglement entropy of our monitored QPC, that we write as~\cite{alba2022noninteracting,alba2022unbounded,soares2025entanglement}:
\begin{equation}\label{eq_qp}
\overline{S}_{\rm{L,R}}(t) = \int_{-\pi}^{\pi} \frac{dk}{2\pi} \min( |v_k| t, L) \, {\rm{h}}(\langle \hat{n}_k(t) \rangle) ,
\end{equation}
where ${\rm{h}}(\expval{\hat n_k})$ accounts for particle loss during the evolution and $\expval{\hat n_k}=\expval{\hat c_k^\dag \hat c_k}$. This quasi-particle picture qualitatively reproduces all the dynamical regimes of the entanglement entropy from our simulations~\cite{supplemental} and allows to obtain analytical insights on the intermediate and long-time scaling behavior. In particular it provides a natural framework to explain the common scaling between entanglement and total particle number. Indeed, for intermediate-long times one can show that the momentum distribution of quasiparticles $\expval{\hat n_k}$ acquires a Gaussian shape~\cite{supplemental}, 
\begin{equation}\label{eqn:nk}
\expval{\hat n_k} \propto \langle \hat N_{\rm{tot}}\rangle e^{-k^2/2\sigma(t)}.
\end{equation}
with a height determined by the average particle number $\langle \hat N_{\rm{tot}}\rangle$ and a width $\sigma(t)$ which decreases in time.
The evolution of this profile consistently shrinks it in both height and width, eventually taking the system to the vacuum at long times. Plugging this ansatz in Eq.~\ref{eq_qp} and taking the long-time limit when $\expval{\hat n_k}\ll 1$ one obtains $\overline{S}_{\rm{L,R}} \propto \langle \hat N_{\rm{tot}}\rangle$~\cite{supplemental}. Consequently, the entanglement entropy inherits the power-law behavior emerging from the evolution of the total number of fermions, $\overline{S}_{\rm{L,R}}\sim\langle\hat N_{\rm{tot}}\rangle \sim 1/(Jt)^\alpha$, as well as the final exponential tail. Remarkably, the quasiparticle picture in Eq.~(\ref{eq_qp}) also accounts for the presence of the volume-law regime at short-times. In general, the quasi-particle picture tends to overestimate the value of the entanglement entropy~\cite{supplemental}. Even accounting for the trajectory-averaged entanglement entropy within the quasi-particle framework cannot fully resolve this discrepancy. This discrepancy in the quasi-particle picture is reminiscent of the rescaling usually found in the unitary case, where the logarithmic conformal formula is modified by an effective central charge in the presence of scattering on a defect~\cite{thomas2015entanglement,eisler2012entanglement}. Indeed, by including
in Eq.~(\ref{eq_qp}) the contribution of quasiparticles which are not absorbed (detected) by the QPC as done in Eq.~(\ref{eqn:slope}) leads to a much better quantitative agreement~\cite{supplemental}.
 
\emph{Full counting statistics of charge transport -- } We now connect our findings on the entanglement entropy under monitoring to an experimentally relevant probe in quantum transport settings, namely the Full Counting Statistics (FCS) of the transported charge~\cite{PismaZhETF.58.225,nazarov2009quantum}. Indeed it is well known for non-interacting fermions under unitary dynamics that the generating function of the FCS is related to the entanglement entropy~\cite{klich2009quantum,Calabrese_2012,song2011entanglement,song2012bipartite,thomas2015entanglement,BURMISTROV2017140}, a result that carries over to the monitored case provided Gaussianity is preserved along the quantum trajectories~\cite{poboiko2023theory,moghaddam2023exponential,tirrito2023full,poboiko2024measurement}. 
Specifically, given the cumulant generating function $\chi(\lambda)=\langle{{\mathrm{exp}}(i\lambda\hat N_{\mathrm{L}})}\rangle$ where $\hat N_{\rm{L}}(t)=\sum_{i=1}^L  \hat{c}_{i\rm{L}}^\dag \hat{c}_{i\rm{L}} $ is the total charge on the left lead,
and the cumulants $\mathcal{C}_n = (-i\partial_\lambda)^n \log\chi(\lambda)|_{\lambda=0}$, the entanglement entropy can be approximated up to order $K$ as~\cite{song2012bipartite}:
\begin{equation}\overline{S_{\rm{L,R}}} \approx \sum_{n=1}^{K+1} \alpha_n(K) \overline{\mathcal{C}_n},
\end{equation}
where $\alpha_n= 2\sum_{k=n-1}^K {\mathrm{S}}_1(k,n-1)/(k!\,k)$ with ${\mathrm{S}}_1$ the Stirling number of the first kind for $n$ even and $\alpha_n=0$ for $n$ odd.
We show in Fig.~\ref{fig2_qp}(c) that very few cumulants are needed to well reproduce the behaviour of the entanglement. Considering cumulants up to $\overline{\mathcal C_4}=\overline{\expval{\left(\hat N_{\rm L}-\expval{\hat N_{\rm L}}\right)^4}-3\left(\expval{\hat N_{\rm L}^2}-\expval{\hat N_{\rm L}}^2\right)}$ gives an excellent reproduction of the entanglement dynamics, as seen in the main figure, and even considering only $\overline{\mathcal C_2} = \overline{\expval{\hat N_{\rm L}^2}-\expval{\hat N_{\rm L}}^2}$ we are able to qualitatively reproduce all the features of the entanglement, as shown in the inset.
Accessing the FCS experimentally is much simpler than the entanglement entropy and natural both in mesoscopic physics settings,
where beyond-Gaussian fluctuations in QPCs have been already observed~\cite{gershon2008detection}, as well as with ultracold atoms~\cite{wienand2024emergence}. Although one would still require post-selecting a single quantum trajectory, whose probability a priori is exponentially small in system size and time,  our setting involves only one monitored channel at the QPC reducing the exponential overhead in system size.

\emph{Discussion - }The linear growth of entanglement entropy under monitoring of particle losses at the QPC, its maximum at times $t\sim L$, followed by its slow-decay in time towards the vacuum is reminiscent of the Page curve~\cite{page1993information,page2013time}, describing the evolution of entanglement entropy during the process of black-hole evaporation and emission of Hawking radiation. Recently, analogue toy models to study the Page curve have been proposed in the context of quantum many-body systems where a filled chain is coupled to an empty one and slowly depleted~\cite{kehrein2024page,saha2024generalized,jha2025page,li2025sharp,ganguly2025quantum}. In those cases the entanglement entropy under unitary dynamics displays again non-monotonic behavior. In our monitored QPC an equivalent effect is obtained by unraveling the dissipative evolution due to particle losses into pure-state quantum trajectories. Our results confirm the breakdown of the semiclassical regime, where entanglement is controlled by the current flowing out of the systems as captured by Eq.~(\ref{eqn:slope}) and offer new insights on the long-time regime by providing a new connection between entanglement and charge dynamics. They also highlight how a monitored QPC could provide an experimentally viable platform for quantum simulation of the Page curve.

\emph{Conclusions -- } In this work we have studied the entanglement dynamics across a continuously monitored QPC, where particle losses at the QPC site are detected according to a quantum jump protocol. We have shown that this local, single-site, monitoring completely reshapes the structure of the entanglement production: instead of the characteristic log-scaling in time of the unitary case we observe a linear growth in time, a maximum entanglement displaying volume-law scaling and a slow decay towards zero. The different dynamical regimes can be all understood with a simple and transparent quasiparticle picture: in particular the linear-in-time growth arise from an emergent bias voltage drop due to the losses of particles at the QPC, while the long-time dynamics of the entanglement entropy is directly tied to the depletion dynamics of the total charge. Using the connection between entanglement entropy and full counting statistics we have connected our results to experimentally relevant quantities and shown that, quite remarkably, only a few cumulants are needed to well describe the entanglement entropy. Finally, we have connected our findings to studies of the Page curve for entanglement entropy. Our work establishes monitored quantum transport as ideal setting where to experimentally investigate the effect of quantum measurements on entanglement entropy, relevant for both mesoscopic circuits as well as cold-atoms quantum transport simulators. A natural direction for future studies concerns the role of many-body interactions on entanglement in these monitored quantum impurity models.

\emph{Acknowledgments} We acknowledge useful discussions with S. Murciano and L. Capizzi. Authors acknowledge computational resources on the Colle\'ge de France IPH cluster. We acknowledge funding from the European Research Council (ERC) under the European Union’s Horizon 2020 research and innovation programme (Grant agreement No. 101002955 — CONQUER).

\bibliography{biblio}
\bibliographystyle{apsrev4-2}

\end{document}


\date{\today}
	
	\title{Supplemental Material to `Entanglement Dynamics across a Monitored Quantum Point Contact'}
	
	\author{Anna Delmonte}
	\affiliation{JEIP, UAR 3573 CNRS, Coll\`ege de France, PSL Research University, 11 Place Marcelin Berthelot, 75321 Paris Cedex 05, France}
	\author{Marco Schir\`o}
	\affiliation{JEIP, UAR 3573 CNRS, Coll\`ege de France, PSL Research University, 11 Place Marcelin Berthelot, 75321 Paris Cedex 05, France}

\maketitle

\onecolumngrid

\renewcommand{\thefigure}{S\arabic{figure}}
\renewcommand*{\citenumfont}[1]{S#1}
\renewcommand*{\bibnumfmt}[1]{[S#1]}

\newcounter{ssection}
\stepcounter{ssection}

\setcounter{table}{0}
\setcounter{page}{1}
\setcounter{figure}{0}
\setcounter{equation}{0}

\makeatletter
\renewcommand{\theequation}{S\arabic{equation}}
\tableofcontents

\section{Numerical Solution of Quantum Jumps Dynamics}

For completeness in this Section, we provide the basic methodology employed to solve the Quantum Jump dynamics introduced in the main text. Specifically, we encode our dynamics in the correlation matrix $C_{i\alpha,j\beta}=\expval{\hat c^\dag_{i\alpha}\hat c_{j\beta}}$. The restriction to the matrix $C$ can only be achieved since the anomalous correlations $\expval{\hat c^\dag_{i\alpha}\hat c^\dag_{j\beta}}$ are always zero in this problem, due to piecewise charge conservation along trajectories. The evolution can be simulated with the following steps.
\begin{itemize}
    \item At the beginning of the step $\delta t$ the probability of having a jump is calculated: 
    \begin{equation}
        p_1 = 2\gamma\delta tC_{L{\rm{L}},L{\rm{L}}}.
    \end{equation}
    \item With a probability $p_1$ the system undergoes a jump. The correlation matrix is then evolved as:
    \begin{equation}
        C_{i\alpha,j\beta}^J = C_{i\alpha,j\beta}-\frac{C_{i\alpha,L\rm{L}}C_{L{\rm L},j\beta}}{C_{L{\rm L},L\rm L}}
    \end{equation}
    \item With a probability $p_0=1-p_1$, the system evolves smoothly along the $\delta t$ step according to the non-linear and non-Hermitian master equation
    \begin{equation}
        \dot{\hat\rho} = -i[\hat H,\hat\rho] -\gamma\left\{\hat c^\dag_{L\rm L}\hat c_{L\rm L},\hat\rho\right\} +2\gamma\hat\rho\Tr{\hat\rho\hat c_{L\rm L}^\dag\hat c_{L\rm L}}.
    \end{equation}
    Considering $\hat H = \sum_{ij}\Vec{\hat c}^\dag h\Vec{\hat c}$ with $\Vec{\hat c}=(\hat c_{1\rm L},...,\hat c_{L\rm L},\hat c_{1\rm R},...,\hat c_{L\rm R})^T$, the above equation reads in terms of the $C$-matrix (using Wick's theorem):
    \begin{equation}
        \dot C = i\left[h,C\right] +\,C\,D\, C -\frac{1}{2}(CD+DC).
    \end{equation}
    with $D_{ij}=2\gamma\delta_{iL}\delta_{jL}$.
    The differential equation is then solved over the time interval $\delta t$. Runge-Kutta 4 allows us to choose a bigger value for $\delta t$.
\end{itemize}

\section{Additional Results on the Entanglement Entropy}

In this Section, we provide further results on the dynamics of the entanglement entropy in our monitored QPC. 

\begin{figure}
\centering 
\includegraphics[width=\linewidth]{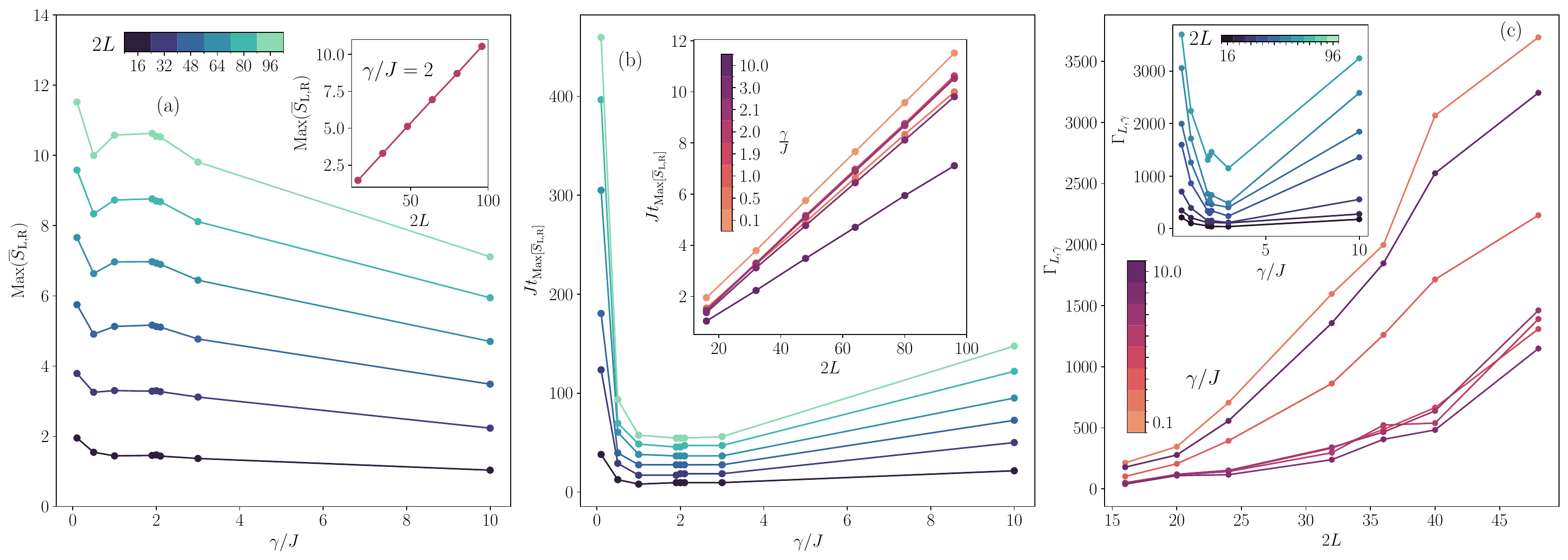}  
    \caption{(a)Maximum of the entanglement entropy versus measurement-strength for various system sizes. In the inset Volume-law at maximum for $\gamma/J=2$. (b) Zeno effect on the time at which the maximum of the entanglement entropy is reached, with the same legend as the other plots. In the inset: linear increase in system size of the time at which the maximum is reached. (c) Behavior of the decay rate of the exponential regime extracted from a fit for different sizes. In the inset, same decay rate with respect to measurement strength.  }
    \label{fig:ent_props}
\end{figure}

We first consider the short-time linear growth regime and discuss the dependence of the maximum the entanglement entropy, $\mbox{Max}(\overline{S}_{\rm{L,R}})$, from the monitoring rate and system size. Fig.~\ref{fig:ent_props} (a) shows that $\mbox{max}(\overline{S}_{\rm{L,R}})$ is non-monotonous in $\gamma/J$: it initially decreases, has a slight increase around $\gamma/J=2$ and decreases for increasing measurement strengths, going to zero in the infinite measurement limit.  As mentioned in the main text, the entanglement entropy displays a clear volume law at maximum, shown in the inset of panel (a) for $\gamma/J=2$. 

We then consider the time scale at which the maximum entanglement entropy is reached, that we denote as $t_{\mbox{Max}(\overline{S}_{\rm{L,R}})}$
and study its dependence from the monitoring rate and system size. In Fig.~\ref{fig:ent_props} (b) we see a clear non-monotonic behavior emerging with $\gamma$, which can be understood as follows. At weak monitoring the dynamics is almost unitary and so the maximum is pushed at times $~1/\gamma$, while for strong monitoring the system displays the Quantum Zeno effect, i.e. the dynamics froze due to repeated measurements and the system takes longer to reach the volume law regime. Concerning the system size dependence of $t_{\mbox{Max}(\overline{S}_{\rm{L,R}})}$ we can see in 
the inset of Fig.~\ref{fig:ent_props} (b) that this time scale increases linearly with system size, indicating that in the thermodynamic limit the system does not fully deplete. This is shown in the inset for several values of $\gamma/J$.

Another timescale which is important to analyze is the exponential decay time at finite sizes $\Gamma_{L,\gamma}$, appearing in the exponential fit of Fig. 2 (d) in the main text. Its behavior is shown in Fig.~\ref{fig:ent_props}(c), as captured by exponential fits on the simulation data.
In general, it depends both on the measurement strength $\gamma/J$ and on the size of the system $L$. Panel (c) shows its dependence on the measurement strength, which resembles in shape Fig.~\ref{fig:ent_props} (c) and is suggestive of a Zeno behavior for the onset of the exponential decay to the vacuum. The inset of panel (c) shows its dependence on system size instead, and proves that its value increases for bigger systems. This indicates once more that the exponential decay does not contribute to the dynamics in the thermodynamic limit.

\section{Charge dynamics across the QPC}

\begin{figure}
\centering 
\includegraphics[width=\linewidth]{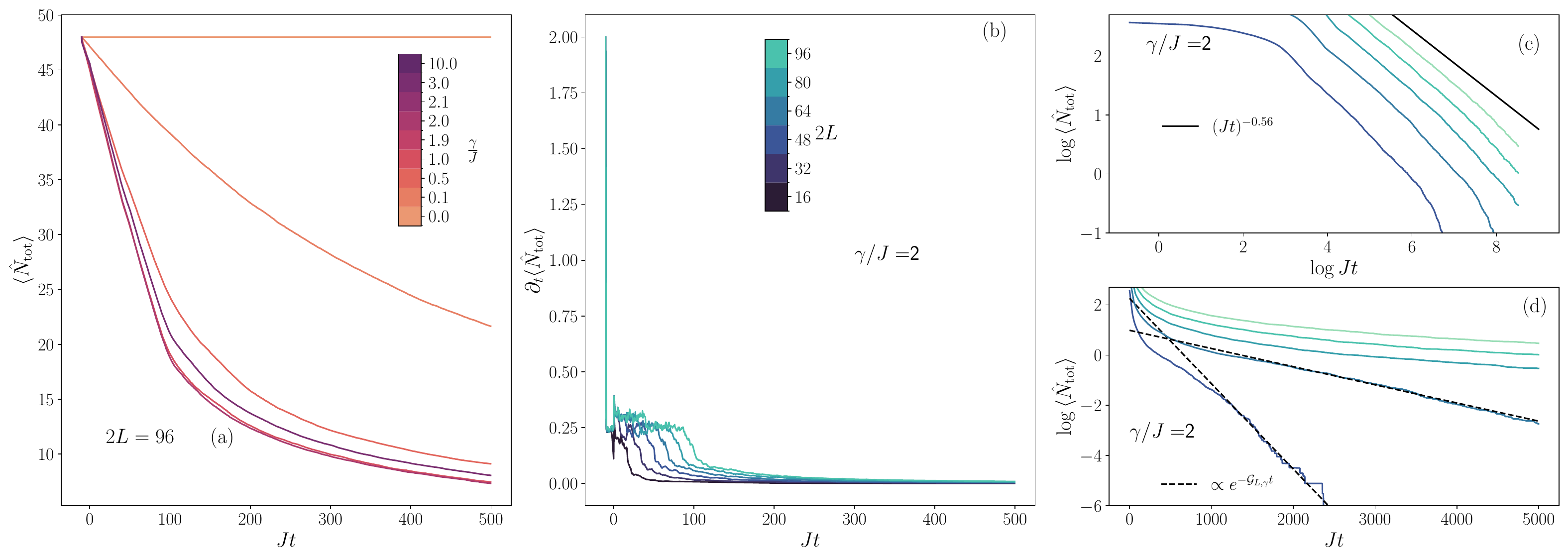} 
    \caption{(a)Dynamics of the total number of particles for several measurement strengths. (b) Time derivative of the total number of particles. (c) Polynomial regime for intermediate times. (d) Exponential regime for long times in finite systems.}
    \label{fig:dens_props}
\end{figure}

In this Section we summarize few results on the dynamics of the average state, which evolves according to the Lindblad master equation
\begin{align}
\partial_t\hat\rho=-i[\hat H,\hat\rho]+\mathcal D[\hat M]\hat\rho
\end{align}
with $\hat H=\hat H_{\rm L}+\hat H_{\rm R}+\hat H_{\rm qpc}$ given in the main text, $\mathcal D[o]\bullet = \hat o\bullet\hat o^\dag-1/2\{\hat o^\dag\hat o,\bullet\}$ and jump operator $\hat M=\sqrt{2\gamma}\hat c_{L \rm L}$. In the present context, this problem describes free fermions with a localized loss at the QPC and displays slow depletion dynamics which has been studied in detail~\cite{froml2019fluctuation,froml2020ultracold}. The equations of motion for the total particle number (total charge) in each lead,
$\hat N_{\rm L}=\sum_{i=1}^L \hat{c}^{\dagger}_{i\rm L}c_{i\rm L}$ and $\hat N_{\rm R}=\sum_{i=1}^{L} \hat{c}^{\dagger}_{i\rm R}c_{i\rm R}$ read
\begin{align}
\partial_t\langle \hat N_{\rm L}\rangle&=\langle  \hat{\mathcal{J}}_{\rm qpc}\rangle-2\gamma\langle n_ {L\rm L}\rangle\\
\partial_t\langle \hat N_{\rm R}\rangle&=-\langle  \hat{\mathcal{J}}_{\rm qpc}\rangle
\end{align}
where $\langle  \hat{\mathcal{J}}_{\rm qpc}\rangle=iJf(t)\langle \hat c^{\dagger}_{L\rm L}\hat c_{1\rm R}-\mbox{h.c.}\rangle$ is the current across the QPC. We now focus on the dynamics of the total charge $\langle \hat N_{\rm tot}\rangle $, that we show in Fig.~\ref{fig:dens_props}. From the equations above we see that the depletion dynamics is controlled by the density of particles at the QPC, $\partial_t\langle \hat N_{\rm tot}\rangle=-2\gamma\langle n_{L,\rm L}\rangle$. 
In Fig.~\ref{fig:dens_props} (a) we plot the dynamics of $\langle \hat N_{\rm tot}\rangle $ for increasing $\gamma/J$, showing a monotonic decay towards the vacuum. Interestingly, also here we see a non-trivial dependence on the monitoring rate, that first speeds up the depletion but then slows it down for $\gamma/J\gg 1$. The initial short-time dynamics is dominated by a linear decrease in the number of particles, as can be seen from the plateau in the time derivative of the total number of particles in panel (b). 
The length of the plateau increases with system size, while its value is independent on $L$. From the Lindblad dynamics we see that this regime corresponds to particle density at the impurity (QPC) site getting closer to a quasi-steady-state, after the initial short-time drop.
The second dynamical regime features a universal polynomial decrease as $(Jt)^{-0.56}$, as shown  in Fig.~\ref{fig:dens_props} (c).
The universal power is indicated in the Figure with a black solid line. The last dynamical regime, in which only finite-size system enter, is an exponential one, in which the decay rate $\mathcal{G}_{L,\gamma}$ depends on $\gamma/J$ and the system size. In panel (d) we show this regime through a numerical fit indicated by the dashed lines. 
Finally, we discussed the charge imbalance $\langle \Delta\hat N \rangle$ whose dynamics from the equations above reads
\begin{align}
\partial_t\langle \hat N_{\rm R}-\hat N_{\rm L}\rangle&=-2\langle  \hat{\mathcal{J}}_{\rm qpc}\rangle-2\gamma\langle n_{L\rm L}\rangle\\
\end{align}
from which we conclude that the charge imbalance can reach a steady-state, provided the current across the QPC compensates the local loss at the QPC site, i.e. $\langle  \hat{\mathcal{J}}_{\rm qpc}\rangle=-\gamma \langle n_{L\rm L}\rangle$.

\section{The quasi-particle picture for entanglement entropy under monitoring}

\begin{figure}
\centering 
\includegraphics[width=0.3\linewidth]{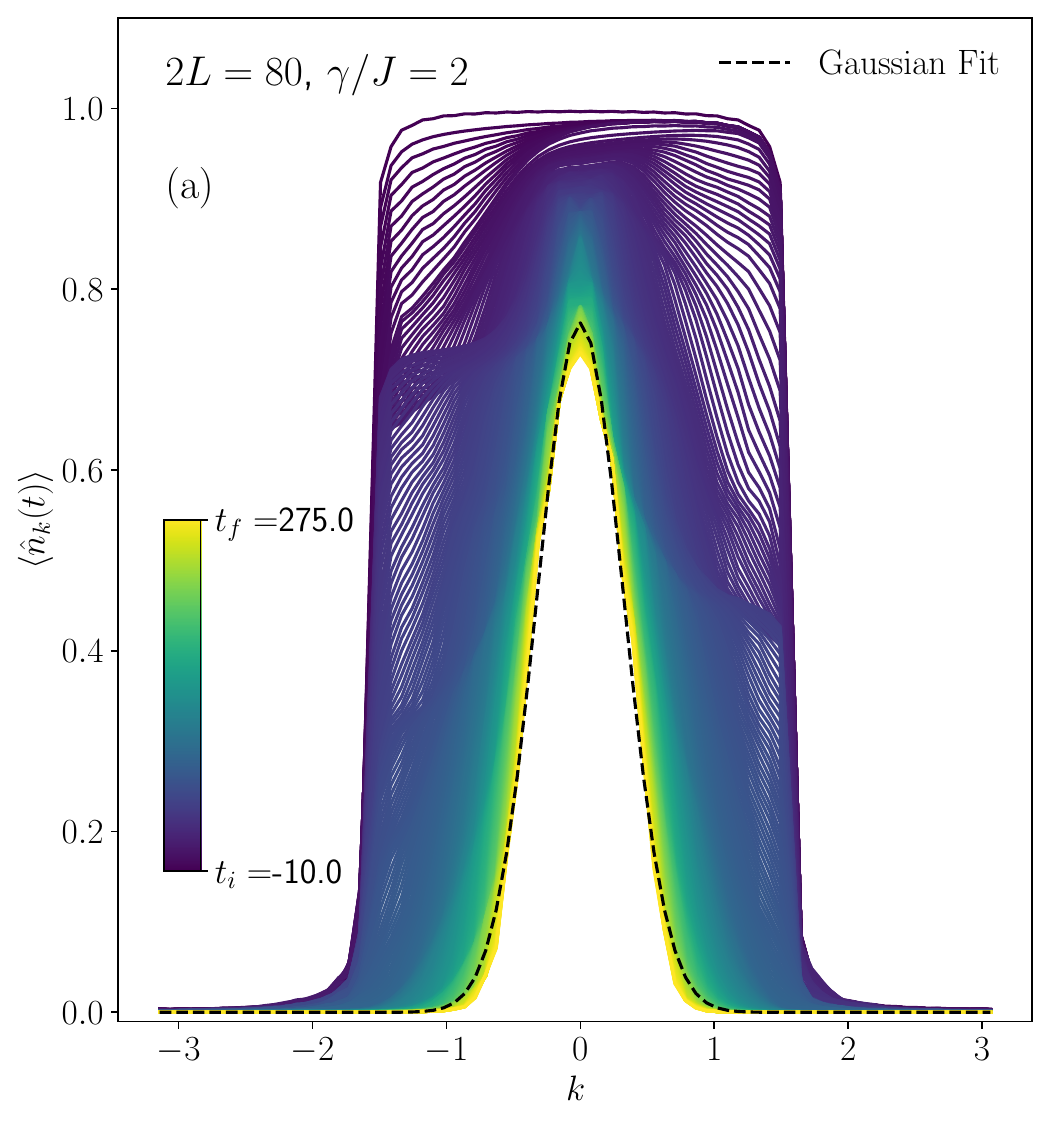} 
\includegraphics[width=0.3\linewidth]{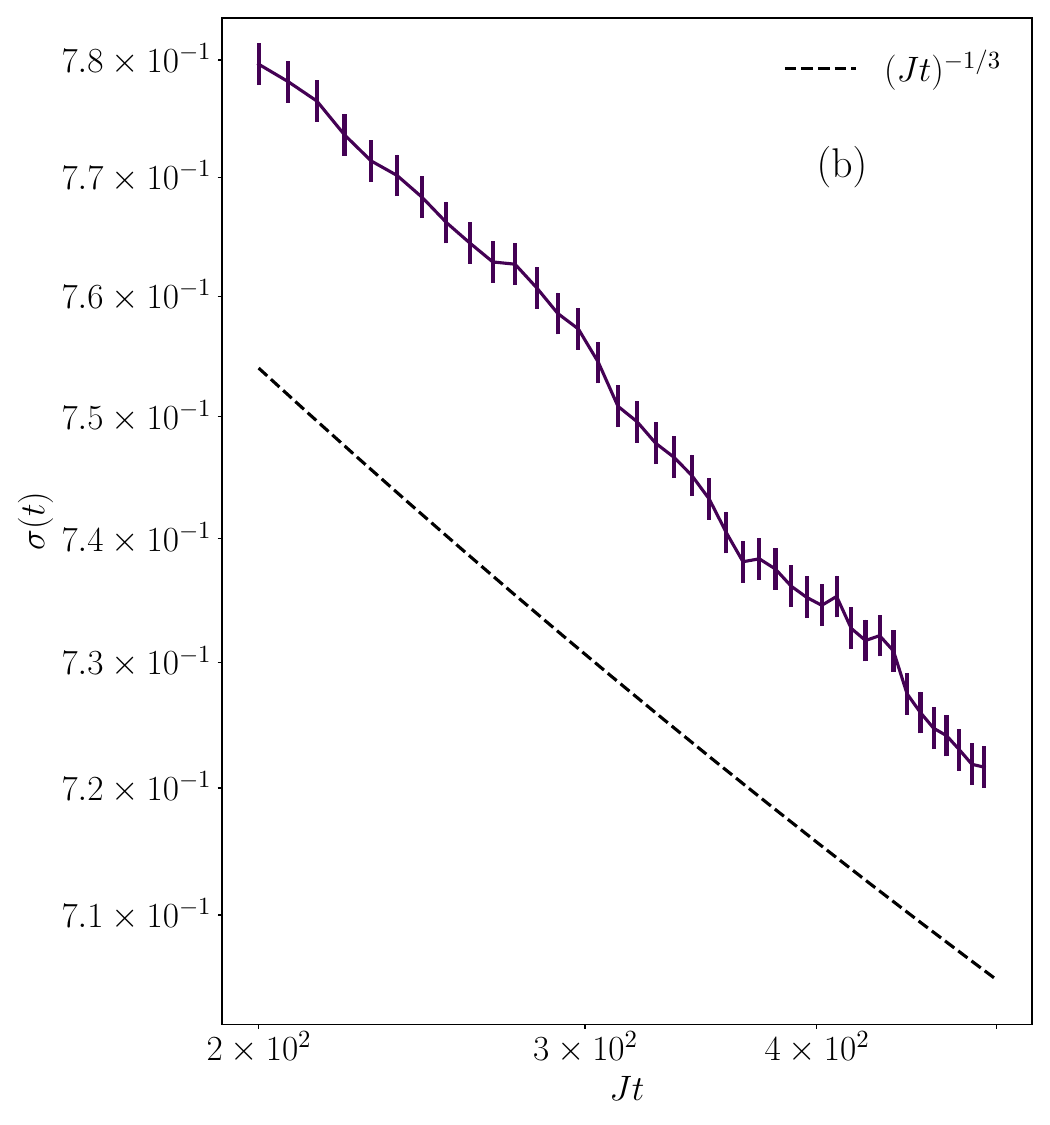}
\includegraphics[width=0.35\linewidth]{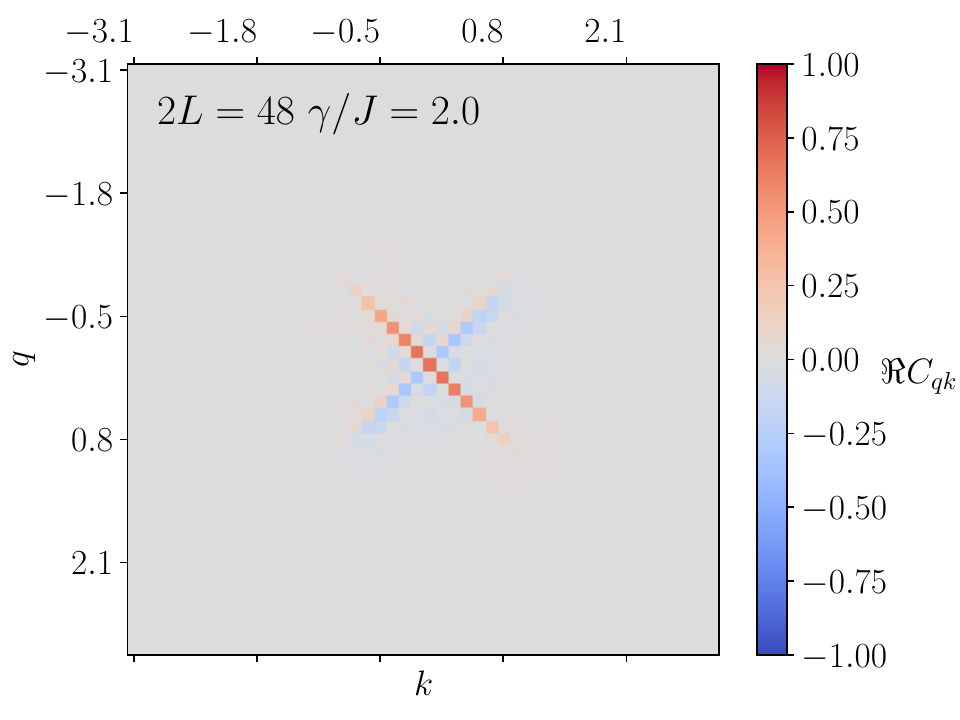}
    \caption{(a) Time evolution of the density profile in momentum space. Time follows the color legend. The Gaussian fit at long times corresponds to the black dashed line. (b) Time evolution of the width of the density profile. (c) Full structure of the density matrix at intermediate times.}
    \label{fig:mom_sp}
\end{figure}

In this Section we provide further details on the quasiparticle picture for the entanglement entropy of the monitored state, introduced in the main text. The basic idea is that while at short-times the entanglement is carried by ballistically propagating quasiparticles, once $t\sim L$ is reached the entanglement entropy decay is completely controlled by the depletion of quasiparticles due to the monitoring process. This leads to the equation
\begin{align}\label{eq_supp_qp}
    S_{\rm{L,R}}(t) &= \int_{-\pi}^{\pi} \frac{dk}{2\pi} \min( |v_k| t, L) \, {\rm{h}}(\langle \hat{n}_k(t) \rangle)
\end{align}
where $h(x)=$ and $\langle \hat{n}_k(t) \rangle$ describes the dynamics of the quasiparticle number with momentum $k$ under the dissipative dynamics. The latter can be obtained from the correlation matrix as
\begin{equation}
    \expval{\hat n_k}=\expval{\hat c_k^\dag c_k} = \frac{1}{2L}\sum_{j,j'}\sum_{\alpha,\beta}e^{-ik\,{\rm d}(j\alpha,j'\beta)}\expval{\hat c_{j\alpha}^\dag c_{j'\beta}}.
\end{equation}
with ${\rm d}(i\alpha,j\beta)=i-j$ if $\alpha=\beta$ or ${\rm d}(i{\rm L},j'{\rm R})=(i-(j+L))$, ${\rm d}(i{\rm R},j'{\rm L})=(L+i-j)$.
The result from simulations is shown in Fig.~\ref{fig:mom_sp}, in which the evolution from the initial state (half-filling) to an intermediate time $Jt_f=275$ is shown. The figure shows that the profile at late times can be fitted with a Gaussian-like function 
\begin{equation}
    \expval{\hat n_k(t)}=c_0(t)e^{-k^2/2\sigma(t)}.
\end{equation}
The height of the Gaussian function is set by the number of particles as $\sum_k\expval{\hat n_k} = \langle \hat N_{\rm{tot}}\rangle$ which can be then set as $c_0(t)\sim\frac{\langle \hat N_{\rm{tot}}\rangle}{\sqrt{2\pi\sigma(t)}}$. Given that $\langle \hat N_{\rm{tot}}\rangle\sim t^{-\alpha}$ with $\alpha\sim0.56$ at intermediate times, and considering the power-law decay of the variance in Fig.~\ref{fig:mom_sp}(b), we notice that the Gaussian functions decreases in height and width, until it gets completely suppressed and $\expval{\hat n_k}=0$ $\forall k$.

Once the Ansatz is built, it can be plugged into the entanglement entropy formula from the quasi-particle picture. In particular, given that $\expval{\hat n_k}\ll1$ we have $\rm{h}(\expval{\hat n_k})\sim\expval{\hat n_k}(1-\log(\expval{\hat n_k}))$. Moreover, at intermediate times and for occupied momenta, we can approximate $-\log\left(\frac{N_{\rm{tot}}(t)}{\sqrt{2\pi\sigma(t)}e^{-k^2/2/\sigma(t)}}\right)\ll1$, yielding
\begin{align}\label{eq_supp_qp}
    S_{\rm{L,R}}(t) &= L\int_{-\pi}^{\pi}\frac{dk}{2\pi}\frac{\langle \hat N_{\rm{tot}}\rangle}{\sqrt{2\pi\sigma(t)}}e^{-k^2/2\sigma(t)}\sim \langle \hat N_{\rm{tot}}\rangle,
\end{align}
showing the relation between the entanglement and the total number of particle for the quasi-particle picture. We note that the width $\sigma(t)$ drops out of the integral and does not contribute to the dynamics of the entanglement, leading to the same scaling with $\langle \hat N_{\rm tot}\rangle$ as observed numerically.

\begin{figure}
\centering 
\includegraphics[width=\linewidth]{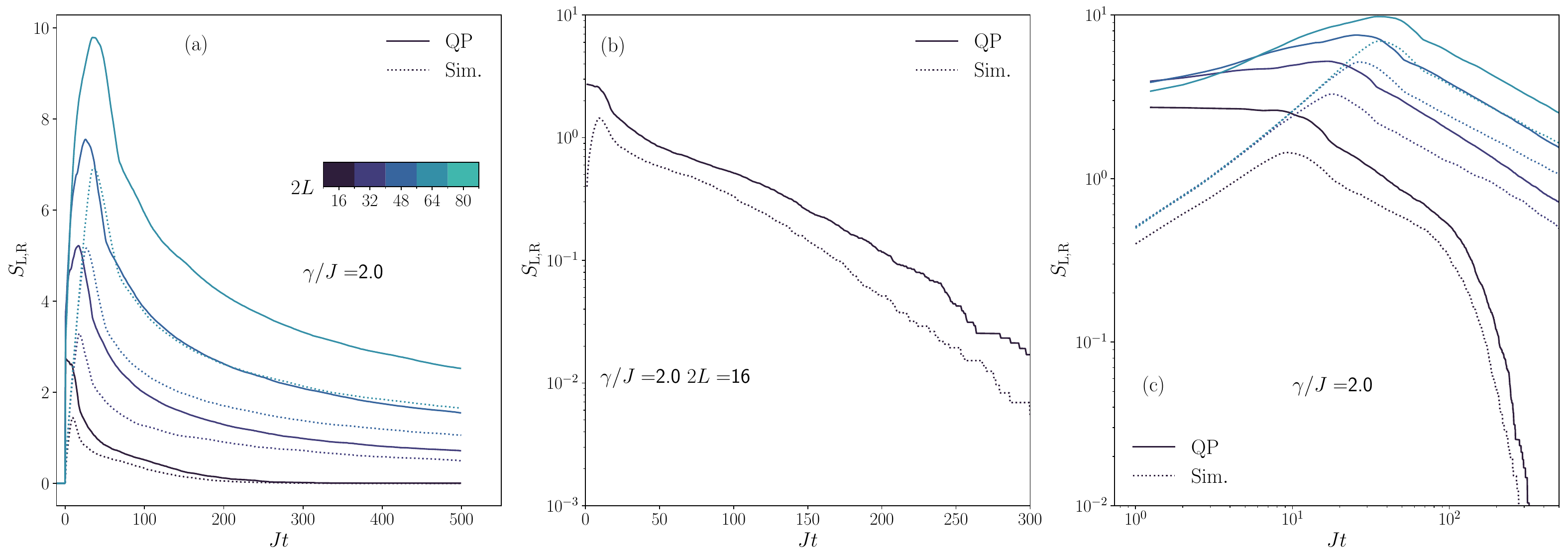} 
    \caption{(a) Comparison of the entanglement entropy obtained from numerical simulations (dotted lines) and the quasi-particle picture (solid lines). (b) The exponential regime of the quasi-particle picture matches simulations. (c) The same power-law regime as in our simulations is also found in the quasi-particle picture.}
    \label{fig:qp_mod}
\end{figure}

The comparison between the quasi-particle picture and the results of our exact simulations is shown in Fig.~\ref{fig:qp_mod}. The quasi-particle picture yields qualitatively the same behaviour of the entanglement entropy as our simulations, but slightly overestimates it, as in panel (a). However all the dynamical features are well captured by the quasi-particle approximation: for long times, the exponential tail has the same behaviour as the simulations, as shown in panel (b); the polynomial regime also arises in the quasi-particle picture and presents the same power as the simulations, as seen in panel (c).

The mismatch in the numerical values of the simulations and the quasi-particle picture is partly due to the fact that we extract $h(\expval{\hat n_k})$ from the Lindblad and not for single trajectories. However, even taking into account $\overline{h(\expval{\hat n_k})}$, the difference in values is not totally filled. We thus have used the values obtained from Lindblad to perform our analytical calculations, which would be otherwise not computable with standard techniques.

So far, we have considered only the diagonal part of the correlation matrix. One can show that, at intermediate times, the correlation matrix develops a pronounced cross-diagonal structure, as shown in Fig.~\ref{fig:mom_sp}(c). This structure reflects the scattering process in which a plane wave with momentum $k$ hits the quantum point contact and is then partially reflected into momentum $-k$, partially transmitted at momentum $k$, and partially absorbed. In the next section, we describe this scattering process and derive the corresponding transmission, reflection, and absorption probabilities. We also briefly discuss how incorporating the scattering process into the entanglement-production framework naturally accounts for the cross-diagonal structure of the correlation matrix.

\section{The scattering problem at QPC}
We treat the scattering problem arising at the quantum point contact (QPC) as a coupling to a measurement device. The last site of the left lead acts as a local impurity where incoming waves from the leads scatter. Considering an incoming plane wave with momentum $k$ originating from the leads, generated by the tight-binding bulk, one can restrict the study to the scattering process at the QPC. This process leads to a mode-dependent reflection probability $R_k$, transmission probability $T_k$, and absorption probability $\eta_k$. The expressions for these quantities can be obtained for a unitary setting, as in Refs.~\cite{thomas2014waiting,thomas2015entanglement}, or in an open setting, as in Ref.~\cite{froml2019fluctuation}; for our specific problem, they read:
\begin{align}
    R_k &= \frac{\gamma^2}{(2J\sin(|k|)+\gamma)^2},\\
    T_k &= \frac{4J^2\sin{k}^2}{(2J\sin(|k|)+\gamma)^2},\\
    \eta_k &= \frac{4J\gamma\sin(|k|)}{(2J\sin(|k|)+\gamma)^2},
\end{align}
with the property $R_k+T_k+\eta_k=1$.

Considering the fact that entanglement is generated by the particles which are not absorbed by the impurity (i.e., detected), we can determine from $\eta_k$ the slope of the initial linear growth of the entanglement entropy, which arises from the binomial process of fermion detection/non-detection at the quantum point contact. In this framework, each mode contributes to the entanglement through ${\rm h}(1-\eta_k)$, summed over all participating momenta~\cite{alba2022unbounded,alba2022noninteracting}. The particle imbalance between the leads determines these momenta. 
Namely:
\begin{equation}
    A_\gamma \sim \int_{k_{\rm L}}^{k_{\rm R}}\,\frac{dk}{2\pi} 2J\abs{\sin(k)}\,{\rm{h}}(1-\eta_k),
\end{equation}
In particular, $k_{\rm{L}} = \frac{\pi}{L}N_{\rm{L}}^*$, and $k_{\rm R} = \frac{\pi}{L}N_{\rm{R}}^*$ are the maximum filled momenta in the two tight-binding bands on the plateau visible in $N_{\rm R}-N_{\rm L}$ in Fig. 3(a). The difference in the number of particles between the two leads can be interpreted as a voltage bias $V \sim 4J \sin(\pi(N_{\rm L} - N_{\rm R}) / 2L)$, given by the differences in Fermi energies between left and right leads. All the momenta (energies) participating in the presence of the bias contribute to the scattering process, which dramatically influences the spread of quasi-particles and thus the growth of entanglement.

\begin{figure}
\centering 
\includegraphics[width=0.33\linewidth]{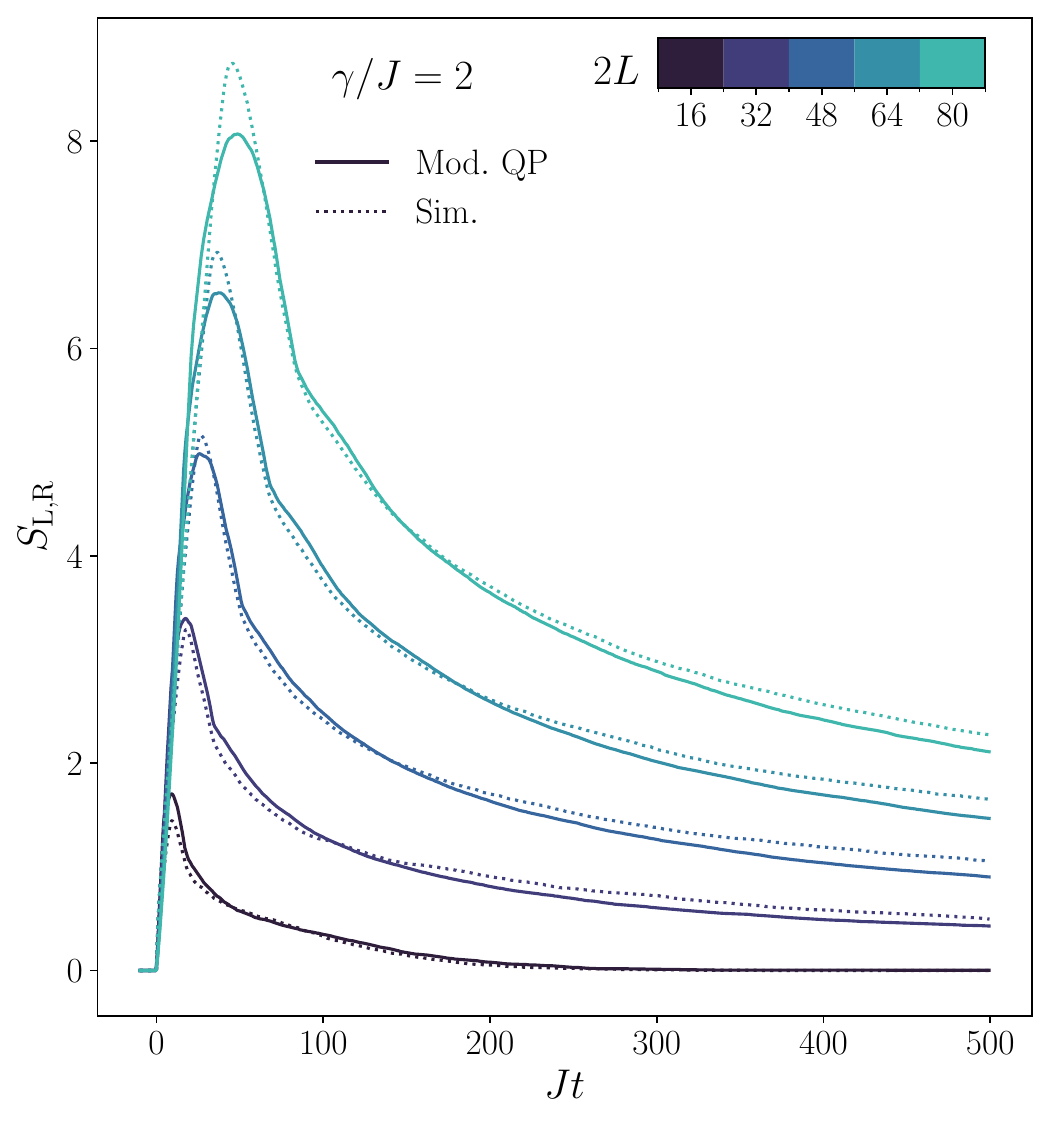} 
    \caption{Comparison of the entanglement entropy obtained from numerical simulations (dotted lines) and the modified quasi-particle picture (solid lines).}
    \label{fig:qpmod2}
\end{figure}

En passant, we note that inserting the same factor ${\rm h}(1-\eta_k)$ to multiply ${\rm h}(\langle \hat n_k \rangle)$ in the quasiparticle picture compensates for the numerical discrepancy between our simulations and the quasiparticle prediction, as shown in Fig.~\ref{fig:qpmod2}. This expression takes into account the cross-diagonal shape of the full correlation matrix through the scattering information. However, this modified quasiparticle picture cannot fully replicate the features of Eq.~\eqref{eq_supp_qp}, as it alters the functional dependence on $k$. A more refined Ansatz must be introduced to reconcile these features and avoid double-counting losses arising from both the absorption coefficients and the time evolution of $\langle \hat n_k \rangle$.

\section{Full-counting statistics}
In the main text we are interested in computing the full-counting statistics of the total number of particles in the left lead $\hat N_{\rm{L}}$. In order to obtain all the cumulants of this operator we can calculate the cumulant generating function:
\begin{equation}
    \chi(\lambda)=\langle{{\mathrm{exp}}(i\lambda\hat N_{\rm L})}\rangle.
\end{equation}
For Gaussian systems, we can exploit the expressions derived in Ref.~\cite{song2012bipartite}, and rewrite the cumulant generating function as
\begin{equation}
    \chi(\lambda) = \det{\left(1+(e^{i\lambda}-1\right)C_{\rm L})},
\end{equation}
where $C_{\rm{L}}$ is simply the correlation matrix of fermions restricted to the left lead: $C_{\rm{L}_{ij}}=\expval{\hat c_{i\rm L}^\dag \hat c_{j\rm L}}$, with $i,j=1,...,L$.

Cumulants are then computed through derivatives of the cumulant generating functional:
\begin{equation}
    \mathcal{C}_n = (-i\partial_\lambda)^n \log\chi(\lambda)|_{\lambda=0}.
\end{equation}

Cumulants are non-linear functions of the state of the system, and thus we need to calculate the cumulant generating function along every single trajectory and then average over all the sampled trajectories. The problem then reduces to compute along trajectories
\begin{equation}
    \log\left(\det{\left(1+(e^{i\lambda}-1\right)C_{\rm{L}})}\right)=\Tr{\log\left(\left(1+(e^{i\lambda}-1\right)C_{\rm L})\right)}.
\end{equation}
This is quickly done numerically in trajectory simulations for $\lambda\in[-\epsilon,\epsilon]$ with $\epsilon\ll1$, and then derivatives can be calculated numerically.

This allows to access directly cumulants and then use them to estimate the entanglement entropy as in the main text in Eq. (6). In particular, Fig. 3c approximates all the cumulants of order $n>4$ to be null and takes the $K\to\infty$ limit in the coefficients.

\bibliography{biblio}